\documentclass[twocolumn,prl,showpacs]{revtex4-1}
\usepackage{graphicx,amsmath,amssymb,xspace}
\usepackage{color}

\begin{document}

\newcommand{\ketbra}[2]{| #1\rangle \langle #2|}
\newcommand{\ket}[1]{| #1 \rangle}
\newcommand{\bra}[1]{\langle #1 |}
\newcommand{\Tr}{\mathrm{Tr}}
\newcommand\F{\mbox{\bf F}}%
\newcommand{\h}{\mathcal{H}}

\newcommand{\PSD}{\textup{PSD}}

\newcommand{\C}{\mathbb{C}}
\newcommand{\X}{\mathcal{X}}
\newcommand{\Y}{\mathcal{Y}}
\newcommand{\Z}{\mathcal{Z}}
\newcommand{\sspan}{\mathrm{span}}
\newcommand{\kb}[1]{\ket{#1} \bra{#1}}
\newcommand{\pos}{D}

\newcommand{\thmref}[1]{\hyperref[#1]{{Theorem~\ref*{#1}}}}
\newcommand{\lemref}[1]{\hyperref[#1]{{Lemma~\ref*{#1}}}}
\newcommand{\corref}[1]{\hyperref[#1]{{Corollary~\ref*{#1}}}}
\newcommand{\eqnref}[1]{\hyperref[#1]{{Equation~(\ref*{#1})}}}
\newcommand{\claimref}[1]{\hyperref[#1]{{Claim~\ref*{#1}}}}
\newcommand{\remarkref}[1]{\hyperref[#1]{{Remark~\ref*{#1}}}}
\newcommand{\propref}[1]{\hyperref[#1]{{Proposition~\ref*{#1}}}}
\newcommand{\factref}[1]{\hyperref[#1]{{Fact~\ref*{#1}}}}
\newcommand{\defref}[1]{\hyperref[#1]{{Definition~\ref*{#1}}}}
\newcommand{\exampleref}[1]{\hyperref[#1]{{Example~\ref*{#1}}}}
\newcommand{\hypref}[1]{\hyperref[#1]{{Hypothesis~\ref*{#1}}}}
\newcommand{\secref}[1]{\hyperref[#1]{{Section~\ref*{#1}}}}
\newcommand{\chapref}[1]{\hyperref[#1]{{Chapter~\ref*{#1}}}}
\newcommand{\apref}[1]{\hyperref[#1]{{Appendix~\ref*{#1}}}}
\newcommand\rank{\mbox{\tt {rank}}\xspace}
\newcommand\prank{\mbox{\tt {rank}$_{\tt psd}$}\xspace}
\newcommand\alice{\mbox{\sf Alice}\xspace}
\newcommand\bob{\mbox{\sf Bob}\xspace}
\newcommand\pr{\mbox{\bf Pr}}
\newcommand\av{\mbox{\bf{\bf E}}}
\newcommand{\pabxy}{(p(ab|xy))}
\newcommand{\calQ}{\mathcal{Q}}
\def\be{\begin{equation}}
\def\ee{\end{equation}}

\newcommand{\vnote}[1]{\textbf{[***#1***]}}
\newcommand{\snote}[1]{\textcolor{blue}{\textbf{(Jamie: #1)}}}
\newcommand{\cnote}[1]{\textcolor{red}{\textbf{(Carlos: #1)}}}
\newcommand{\wnote}[1]{\textbf{[***#1***]}}
\newcommand{\comment}[1]{{}}
\newcommand{\inner}[2]{\langle #1, #2 \rangle}
\newcommand{\NEW}[1]{\textcolor{blue}{#1}}
\newcommand{\CHECK}[1]{\textcolor{red}{#1}}
\newcommand{\NEWA}[1]{\textcolor{red}{#1}}
\newcommand{\red}[1]{\textcolor{red}{#1}}

\title{\vspace{-1cm}Minimum Dimension of a Hilbert Space Needed to Generate a Quantum Correlation}

\author{Jamie Sikora$^{1,2}$, Antonios Varvitsiotis$^{1,2,3}$, and Zhaohui Wei$^{1,2,3,}$}
\email{Email: zhwei@ntu.edu.sg} \affiliation{$^{1}$Centre for
Quantum Technologies, National University of Singapore, Singapore
117543\\$^{2}$MajuLab, CNRS-UNS-NUS-NTU International Joint Research
Unit, UMI 3654, Singapore\\$^{3}$School of Physical and Mathematical
Sciences, Nanyang Technological University, Singapore 637371}

%\date{July 2, 2015}

\begin{abstract}
Consider a two-party correlation that can be generated  by
performing local measurements on a bipartite  quantum system. A
question of fundamental importance is to understand how many
resources, which we quantify by the dimension of the underlying
quantum system, are needed to reproduce this correlation. In this
Letter, we identify an easy-to-compute lower bound on the smallest
Hilbert space dimension needed to generate a given two-party quantum
correlation. We show that our bound is tight on many well-known
correlations and discuss how it can rule out correlations of having
a finite-dimensional quantum representation. We show that our bound
is multiplicative under product correlations and also that it can
witness the non-convexity of certain restricted-dimensional quantum
correlations.

\end{abstract}
%\pacs{03.65.Aa, 03.65.Ud, 03.65.Wj}

\maketitle

In what ranks as one of the most important achievements  of modern
physics, {it was shown {by John Bell} in 1964 %by means of Bell's inequality
that some correlations generated within the framework of quantum
mechanics {can} be {\em nonlocal}, in the sense that
%sometimes
the statistics generated by quantum mechanics cannot  always be
reproduced by a  local hidden-variable model~\cite{Bell64,Bell66}}.
{Over the last $40$ years, there have been significant efforts in
trying to verify this fact experimentally.}
 The first such
experimental data~\cite{FC72} was published in $1972$ and {this
remains an active area of research}~\cite{HBD+15}. Moreover, as a
{central} concept in {quantum physics and quantum information
theory}, fully understanding quantum {entanglement and} nonlocality
still remains a very interesting and important problem with
far-reaching applications. Indeed, profound relationships between
quantum nonlocality and other fundamental quantum concepts or
phenomena such as entanglement {measures}~\cite{Werner89,ADGL02},
entanglement distillation~\cite{SCA08,Acin01}, and
teleportation~\cite{CABV13} have been {identified}.
%found
Meanwhile, {for}
%in
 many
tasks, e.g. in cryptography~\cite{Ekert91,SCK14}, it has been
realized that due to quantum nonlocality, quantum strategies enjoy
remarkable advantages over {their classical counterparts}.
%classical ones.
\comment{Applications range from
cryptographic tasks such as key distribution~\cite{Ekert} and
device-independent cryptography~\cite{Yao} to entanglement
distillation~\cite{Acin} to gaining a better understanding of
quantum entanglement itself~\cite{Werner, Masanes, Acin}.}

However, {even though quantum nonlocal effects can lead to
interesting and often surprising advantages in some applications,
this does not paint the full picture. After all, {for practical
applications,} it is just as important to {understand} the amount of
quantum resources required for these advantages to manifest. For
instance, if there is an exponential blowup in the amount of
resources required, then whatever advantage {gained} by employing
quantum mechanics may not be useful in practice. Quantifying the
amount of quantum resources needed to perform a certain task is the
central focus of this Letter.

%In this Letter,
We study quantum nonlocality from the viewpoint of two-party quantum
correlations that arise from a Bell experiment. A two-party Bell
experiment is performed between  two
%experimentalists,
{parties}, Alice and Bob, whose labs are set up in separate
locations. Alice (resp. Bob) has in her possession a measurement
apparatus whose possible settings are labelled by the elements of a
finite set $X$ (resp.  $Y$) and the possible measurement outcomes
are labelled by a finite set $A$ (resp. $B$). After repeating the
experiment many times, Alice and Bob calculate the joint conditional
probabilities $p(ab|xy)$, i.e., the probability that upon selecting
measurement settings $(x,y)\in X\times Y$ they get outcomes
${(a,b)\in A\times B}$. {The collection of all joint conditional
probabilities is arranged in a vector  $p=\pabxy$ of length
${|A\times B\times X\times Y|}$ which we call a {\em correlation}}.

Given a Bell experiment  as described above, a natural problem is to
characterize the correlations that can arise with respect to various
physical models. The set of correlations generated by a local-hidden
variable model forms a convex polytope and its elements are called
{\em local correlations}.  A correlation  $p = \pabxy$ is called {\em
quantum} if it can be generated by performing local measurements on
a shared quantum system which is prepared in a state independent of
the measurement choices. Formally, $p=\pabxy$ is quantum if there exists a quantum state
 $\rho$ acting on the Hilbert space $\C^d \otimes \C^d$ and local {positive-operator valued measures (${\rm
POVM}$s)} ${\{ {M_{xa}} : a \in A \}}$  and $\{ {N_{yb}} : b \in B
\}$ each acting on $\C^d$ such that
\begin{equation}\label{eq:correlation}
p(ab|xy) = \Tr( {(M_{xa} \otimes N_{yb}} )\rho).
\end{equation}
%In the case above,

{For a correlation of the form \eqref{eq:correlation}} we say that
$p$ admits a \emph{$d$-dimensional representation.} {Furthermore, we
denote by $\mathcal{D}(p)$ the minimum integer $d\ge 1$ for which
the correlation  $p$ admits a $d$-dimensional representation.} {Note
that the  case  $\mathcal{D}(p)=1$ corresponds to  local
correlations where Alice and Bob only use private randomness.}

{As we only consider finite-dimensional Hilbert spaces, we can
replace the tensor product structure with commutation relations and
obtain an equivalent definition~\cite{BCP14}.}

{Considering the central role that quantum correlations play in many
applications and the  fact that Hilbert space dimension is a
valuable resource, a natural and fundamental problem is as follows:
Given a quantum correlation $p=\pabxy$, what is the smallest
dimension of a quantum system needed to generate $p$, i.e., what is
$\mathcal{D}(p)$?}

This problem is NP-hard to solve exactly in general~\cite{Stark15},
and limited progress has been reported; see \cite{BCP14} for a
summary of results. One of the most successful  approaches employs
the notion of  {\em dimension witnesses} \cite{BPA+08} (see also
\cite{PV08,WCD08,BBT11}). Furthermore, the framework of dimension
witnesses  has been also used  to derive dimension  lower bounds in
the prepare-and-measure scenario~\cite{GBHA10}.

In the setting {of \cite{BPA+08}}, a $d$-dimensional representation
of a correlation $p=\pabxy$ is defined as a convex combination of
correlations  of the form \eqref{eq:correlation}. Operationally,
this means that the preparations of the quantum states and the POVMs
{depend} on the value of a public random variable, which {they
consider to be} a free resource.

The assumption  of free public randomness implies that the set of
correlations admitting a $d$-dimensional representation, denoted by
$\mathcal{Q}_d$, is convex. A {\em $d$-dimensional witness} is
defined as a hyperplane $H$  that contains $\mathcal{Q}_d$ in one of
its halfspaces. Consequently, for any correlation $p$  that lies
{strictly} in the opposite halfspace, $H$ witnesses that  $p\not \in
\mathcal{Q}_d.$ Note that since $\mathcal{Q}_d$ is convex  such a
hyperplane exists for any $p\not \in \mathcal{Q}_d$. On the negative
side, finding such a hyperplane (for a given  correlation and a
fixed  $d\ge 1$) is a challenging task.

On the other hand, if public randomness is not {a free resource,
i.e., it must be embedded into the entangled state $\ket{\psi}$},
the set of quantum correlations admitting a
 $d$-dimensional representation  (as defined in \eqref{eq:correlation}) is  not always  convex~\cite{DW15}. The  lack of convexity
in this setting suggests that  the problem of lower bounding the
size of the {quantum} system needed to generate a correlation is
more complicated.
   In particular, the approach of using  separating
hyperplanes is no longer applicable.
Nevertheless, this
is a realistic and interesting setting,  %that deserves  to be characterized,
 e.g., when public randomness is    not available,
or when we need to compare the resources required by a classical
scheme and those by a pure quantum scheme to generate a given
correlation.

In this Letter, for the case that public randomness is not a free
resource, we give an \emph{easy-to-compute} lower bound on
$\mathcal{D}(p)$ which only depends on the values of the joint
conditional probabilities $p(ab|xy)$. To derive the bound, we {use
an approach that combines} a novel geometric characterization for
the set of quantum correlations given in \cite{VS15} with techniques
that were recently introduced to lower bound the {Positive
Semidefinite Rank} {(see (\ref{eq:psdrank}) for a definition)} of an
entrywise nonnegative matrix~\cite{LWdW14}, {a fundamental quantity
in both mathematical optimization and quantum communication theory
\cite{FMP+12,GPT13}.} {We then apply our lower bound to show that it
is tight on many {well-known} correlations. Afterwards, we also
detail various other applications.}

{{\em Deriving our lower bound.---}}{The first ingredient in proving
our lower bound on the Hilbert space dimension} relies on the fact
that, without loss of generality, we can assume Alice and Bob share
a pure state on the Hilbert space $\C^d \otimes \C^d$. To argue
this, suppose $p=\pabxy$ is generated by a mixed state $\rho$ acting
on $\C^d \otimes \C^d$. Consider its purification $\ket{\psi} \in
\C^d\otimes \C^d \otimes \Z$, then look at its Schmidt decomposition
${\ket{\psi} := \sum_{i=1}^{d} \lambda_i \ket{a_i}_{\C^d}
\ket{b_i}_{\C^d \otimes \Z}}$, where we allow $\lambda_i = 0$ in the
Schmidt decomposition for convenience. Note that since the first
subsystem is $d$-dimensional we have $d$ terms in the Schmidt
decomposition. Consider the maps  ${U := \sum_{j=1}^{d} \ket{j}
\bra{a_j}}$ and ${V := \sum_{j=1}^{d} \ket{j} \bra{b_j}}$ and define
the  pure quantum state $\ket{\psi'}:=(U \otimes V)\ket{\psi} \in
\C^d \otimes \C^d$, returning to the original Hilbert spaces. By
adjusting  the measurement operators  using $U$ and $V$ we can
construct a $d$-dimensional representation for $p$ using the pure
state $\ket{\psi'}\in \C^d \otimes \C^d$. A similar proof shows that
Alice and Bob's quantum systems can be of the same dimension (being
the minimum dimension of the original two~systems).

The second ingredient in proving our lower bound is a recent
characterization for the  correlations that  admit  a
$d$-dimensional representation with a pure quantum state.
{Specifically,} it was shown in \cite{VS15} that a correlation $p =
\pabxy$ is generated by a pure quantum state $\ket{\psi} \in \C^d
\otimes \C^d$ if and only if there exist $d \times d$ Hermitian
positive semidefinite matrices {$\{ E_{xa} : a \in A, x \in X \}$
and ${\{ F_{yb} : b \in B, y \in Y \}}$} satisfying the following
conditions:
\begin{eqnarray}
p(ab|xy) & = & \Tr(E_{xa} F_{yb}), \text{ for all } a,b,x,y, \label{condition:1} \\
\sum_{a \in A} E_{xa} & = & \sum_{b \in B} F_{yb}, \text{ for all } x,y. \label{condition:2}
\end{eqnarray}

{Combining this}  with the fact that we can assume that a
correlation is generated by a pure state, we have that for a quantum
correlation ${p =\pabxy}$, $\mathcal{D}(p)$ is equal to the smallest
integer $d\ge 1$ for which there exist  $d \times d$ Hermitian
positive semidefinite matrices $\{ E_{xa} : a \in A, x \in X \}$ and
$\{ F_{yb} : b \in B, y \in Y \}$ satisfying (\ref{condition:1})
and~(\ref{condition:2}).

{We now have all the necessary ingredients to derive our lower bound
on $\mathcal{D}(p)$. For the remainder of this section fix a
correlation $p=\pabxy$, set $d:=\mathcal{D}(p)$ and let $\{ E_{xa} :
a \in A, x \in X \}$ and $\{ F_{yb} : b \in B, y \in Y \}$ be two
families of $d \times d$ matrices satisfying \eqref{condition:1} and
\eqref{condition:2}.} Notice that $\sum_{a } E_{xa}$ has full rank
for any $x$ (otherwise, by restricting on its support, we can
construct a new family of matrices of size strictly less than $d$
satisfying (\ref{condition:1}) and (\ref{condition:2}), {which
contradicts} the minimality of $d$). We first create a family of
POVMs by defining the invertible matrix $U$ such that {$U \left(
\sum_{a } E_{xa} \right) U^{\dag} = I_d.$}
Thus, $\{ E'_{xa} := U E_{xa} U^{\dag} : a \in A \}$ is a POVM for any choice of~$x$. Notice %when $p(a,b|x,y) > 0$,
we can write
\begin{equation}
{p(ab|xy)}  = f_{yb} \; \Tr( E'_{xa} F'_{yb}), \label{eqn:6}
\end{equation}
for all $a,b,x,y$, where $F'_{yb} := (U^{-1})^\dag F_{yb}
U^{-1}/f_{yb}$ and $f_{yb}$ is the normalizing factor so that
$F'_{yb}$ is a quantum state. Notice now that $p(ab|xy)/f_{yb}$ is
the probability of outcome $a$ when $F'_{yb}$ is measured with the
POVM {$\{ E'_{xa} : a \in A \}$} when $f_{yb} > 0$. {Recall that
}the  fidelity between two quantum states $\rho$ and $\sigma$ is
defined as $\F(\rho, \sigma) := \| \sqrt{\rho} \sqrt{\sigma} \|_1$.
{Note that} the fidelity can only increase after a measurement
\cite{NC00}, thus we~have
\begin{eqnarray}
& & \!\!\!\!\!\!\!\!\!\! \F (F'_{y_1b_1}, F'_{y_2b_2}) \leq \sum_{a}
\sqrt{\frac{p(ab_1|xy_1)}{f_{y_1b_1}}}
\sqrt{\frac{p(ab_2|xy_2)}{f_{y_2b_2}}} \label{eqn:7}
\end{eqnarray}
for all $x$. Furthermore, we have that $\Tr({\rho}{\sigma}) \leq
\F(\rho,\sigma)^2$, implying\begin{equation} \Tr(F'_{y_1b_1}
F'_{y_2b_2}) \leq \F(F'_{y_1b_1},  F'_{y_2b_2})^2. \label{eqn:8}
\end{equation}
%The fact that
{Since} $p(ab|xy)$ is a probability distribution {for all $x,y$, it
follows from~(\ref{eqn:6})} that $\sum_b f_{yb} = 1$ for all $y$. We
now define the mixed state $\rho_y := \sum_{b} f_{yb} F'_{yb}$ for
all $y$. Since $\sum_{b} F_{yb}$ is independent of $y$ from
$(\ref{condition:2})$, we have that
\begin{equation}
\rho_{y_1} = \rho_{y_2}, \text{ for all } y_1, y_2. \label{eqn:9}
\end{equation}
Since $\rho_y$ is a mixed quantum state over $\C^d$, we have that
\begin{equation}
\Tr(\rho_y^2) \geq \frac{1}{d}, \text{ for all } y.
\label{eqn:10}
\end{equation}
Combining
Equations~(\ref{eqn:7}),~(\ref{eqn:8}),~(\ref{eqn:9}),~(\ref{eqn:10})
{it follows that $d$ is lower bounded by}
%get the following lower bound for $d$
\begin{eqnarray}\label{eq:lbound1}
\!\!\!\!\!\!\!\!\!\!\!\! \max_{y_1, y_2} \! \left[
\displaystyle\sum_{b_1,b_2} \! \min_{x} \! \left( \! \sum_{a}
\sqrt{p(a b_1 | x y_1)} \sqrt{p(a b_2 | x y_2)} \right)^2
\right]^{-1}. \label{eqn:11}
\end{eqnarray}
Note that we could have
%just as easily
transformed the matrices $F_{yb}$ into the measurements instead of
the matrices $E_{xa}$. Repeating the above analysis in this case, we
arrive at
\begin{eqnarray}\label{eq:lbound2}
& &
\!\!\!\!\!\!\!\!\!\!\!\! \max_{x_1,x_2}
\! \left[
\displaystyle\sum_{a_1,a_2}
\! \min_{y}
\! \left(
%{P}_{a_1 *}^{x_1y},{P}_{a_2 *}^{x_2y}
\! \sum_{b} \sqrt{p(a_1 b| x_1 y)} \sqrt{p(a_2 b| x_2 y)}
\right)^2\right]^{-1} \label{eqn:12}
\end{eqnarray}
as another lower bound on $\mathcal{D}(p)$. {We collect these  two
lower bounds on $\mathcal{D}(p)$  in the main theorem of this
Letter,~below.}

\noindent \textbf{Theorem.} {For any %{two-party}
quantum correlation $p$
%=\pabxy}$,
we have  that \be\label{eq:lbound} \mathcal{D}(p)\ge \big\lceil
\max\{ f_1(p),f_2(p)\}\big\rceil, \ee where  $f_1(p)$ and $f_2(p)$
denote  the expressions given in  \eqref{eq:lbound1} and
\eqref{eq:lbound2} respectively, {and $\lceil a\rceil$ is the least
integer $t$ such that $t \geq a$.}

{\em Applications.---} In the rest of this Letter, we illustrate the
usefulness of our lower bound {for various applications}.

{\em Several {well-known} correlations.} {We start by showing  that
the {lower} bound can be tight.}
 Let $A = B = X = Y = \{ 0, 1 \}$,
and consider the quantum correlation given by
\begin{equation}\label{cor:chsh}
p(ab|xy) =
\left\{
\begin{array}{rcl}
(2+\sqrt{2})/8, & \text{ if } & a \oplus b = {xy}, \\
(2-\sqrt{2})/8, & \text{ if } & a \oplus b \neq {xy},
\end{array}
\right.
\end{equation}
where $\oplus$ denotes the logical exclusive OR of two bits. This
correlation corresponds to the optimal strategy for the CHSH
game~\cite{CHSH69} which can be generated using the quantum state
$\frac{1}{\sqrt 2}(\ket{00}+\ket{11}) \in \C^2 \otimes \C^2$.
Applying our lower bound to the above correlation, we obtain
$f_1(p)=2$, which is tight.

We next consider a correlation in the setting $X = Y = \{ 1, 2, 3
\}$, ${A = B = \{ 0, 1 \}^3}$
%$|A| = |B| = \{ 0, 1 \}^3$
 generated using the state
$\frac{1}{2} (\ket{0011}-\ket{0110}-\ket{1001}+\ket{1100}) \in \C^4
\otimes \C^4$ given by
\begin{eqnarray} \label{corr:MSG}
& & \!\!\!\!\!\!\!\!\! p(ab|xy) \! = \! \left\{
\begin{array}{cl}
\! {1/8}, & \text{if } a_y = b_x, \; a \text{ has even parity},  \\
\! \phantom{1/8} & \phantom{\text{if }} \text{\ \ \ and } b \text{ has odd parity}, \\
%\! \phantom{1/8} & \phantom{\text{if }} a_y = b_x, \\
\! 0, & \text{otherwise}. \\
\end{array}
\right.
\end{eqnarray}
This correlation is optimal for the Magic Square
Game~{\cite{Mermin90,Peres90,Aravind02}}. Using~(\ref{eqn:11}), we
can easily show that $f_1(p)=4$, which is again tight.

{In addition to the above examples of extremal correlations, we
would now like to discuss some examples which are non-extremal. We
now discuss correlations in connection to a Bell inequality (Eq.~(5)
in \cite{BPA+08}) where $|X| = |B| = 2$ and $|A| = |Y| = 3$. It was
shown in \cite{BPA+08} that the maximal violations require a
two-qutrit state to achieve. By trying our lower bound on some near
maximally violating correlations (found numerically), we find that
our lower bound yields $2 \pm \epsilon$ for small $\epsilon > 0$.
Thus, after rounding up, it sometimes gives a tight result.
Interestingly, there are some non-local correlations which do not
violate the Bell inequality but our lower bound is strictly greater
than $2$, yielding a tight bound once rounded up. This illustrates
the fact that our bound is independent of any Bell inequalities and
complements the approach of dimension witness.}

{As a last example, we study the I3322 Bell
inequality~\cite{Froissart81}. The maximal value of I3322 is $0.25$
when restricted to using qubit states, and numerical evidence shows
that the maximal violation requires infinite dimensional Hilbert
spaces~\cite{PV10}. When evaluating our lower bound on some
correlations with I3322 value greater than $0.25$, we get values
between $1$ and $2$, which is not tight. Indeed, as the correlations
approach the maximum I3322 value, the probabilities in the numerical
simulations are bounded away from $0$, and thus our lower bound does
not grow large.}

{\em Witnessing the non-convexity of {restricted-dimensional} quantum correlations. }
It is known that the extreme points of the set of quantum
correlations in the $|X| = |Y| = |A| = |B| = 2$ setting can be
generated using {a two-qubit state~\cite{Masanes06}}. It has been
shown numerically that some correlations in this setting require at
least a two-qutrit state to generate~\cite{DW15}, thus implying
%showing
that the set $D_2 := \{
p : \mathcal{D}(p) \leq 2 \}$ is not convex. %In this application,
Using our lower bound we can give an analytical proof of this fact.
Consider the following three deterministic correlations in $D_2$:
\begin{eqnarray*}
p_1(ab|xy) = 1, \ & \text{ if } & (a = 1 \text{ and } b = 1), \; 0 \text{ otherwise}, \\
p_2(ab|xy) = 1,\  & \text{ if } & (a = 0 \text{ and } b = 0), \; 0 \text{ otherwise}, \\
p_3(ab|xy) = 1,\  & \text{ if } & (a \neq x \text{ and } b \neq y), \; 0 \text{ otherwise}.
\end{eqnarray*}
Setting $p = \frac{1}{3} p_1 + \frac{1}{3} p_2 + \frac{1}{3} p_3$ we
have that $f_1(p)=9/4 > 2$. Thus $p \not\in D_2$, witnessing the
non-convexity of $D_2$.

{\em Witnessing non-quantumness.} We now consider a generalization
of the Popescu-Rohrlich box (PR-box)~\cite{PR94,BLM+05} in the
setting  $X=Y=\{0,1\}$, $A=B=\{0,1,\ldots, d-1\}$ given by
\begin{equation}\label{cor:pr}
p(ab|xy) = \left\{
\begin{array}{ccl}
1/d, & \text{ if } & xy = (b - a) \ {\rm mod} \ d, \\
0, & \text{ if } & xy \neq (b - a) \ {\rm mod} \ d.
\end{array}
\right.
\end{equation}
A sufficient condition was derived in~\cite{deVicente15} which
witnesses that $p$ is not quantum {(see also \cite{RTHH14})}. We can
readily verify that {$f_1(p) = + \infty$ yielding an alternative
proof that it has no finite-dimensional quantum representation.}

We proceed to show that  a second family of correlations is not
finite-dimensional quantum. In particular, in the setting ${X = Y =
A = B = \{ 0, 1 \}}$, consider any correlation $p$ which satisfies
\begin{equation}
\!\! p(ab|xy) = 0, \text{ if } (x \vee a = y \vee b)
\end{equation}
when $(x,y) \neq (1,1)$, where $\vee$ denotes the logical OR of two
bits. Such correlations correspond to perfect strategies for the
Fortnow-Feige-Lov\'asz (FFL) Game \cite{FL92,Fortnow89}. It follows
from the computation of the entangled value of this
game~\cite{Watrous04} that such a quantum correlation cannot exist.
By examining the pattern of $0$'s in the correlation, we  can apply
the same argument as before to conclude that there is no
finite-dimensional quantum representation of $p$.

{\em {Multiplicity} of the lower  bound under product correlations.}
For $i \in \{ 1, \ldots, k \}$, consider quantum correlations $p_i$,
on the settings $X_i$, $Y_i$, $A_i$, and $B_i$, respectively. Define
the \emph{product correlation} $p_{1, \ldots, k}$ on $X =
\times_{i=1}^k X_i$, $Y = \times_{i=1}^k Y_i$, $A = \times_{i=1}^k
A_i$, and $B = \times_{i=1}^k B_i$, given by
\begin{equation}
p_{1, \ldots, k} (ab | xy) \\
:= \displaystyle\Pi_{i = 1}^k \; p_i(a_i b_i | x_i y_i).
\label{eqn:prodcorr}
\end{equation}
Clearly, since we can generate $p$ using $k$ separated subsystems,
we have $\mathcal{D}(p_{1, \ldots, k}) \leq \Pi_{i=1}^k \;
\mathcal{D}(p_i)$. We now identify a sufficient condition for this
to hold with equality.

It is straightforward to verify that $f_1$, defined  in
\eqref{eq:lbound1}, multiplies under product correlations, i.e.,
\begin{equation}
f_1(p_{1, \ldots, k}) = \Pi_{i=1}^k \; f_1(p_i).
\end{equation}
Thus, if $f_1(p_i)=\mathcal{D}(p_i)$ for all $i\in \{1,\ldots,k\}$ we get~that
\begin{equation} \mathcal{D}(p_{1, \ldots, k}) = \Pi_{i=1}^k \; \mathcal{D}(p_i).
\end{equation}
Clearly, the same argument holds if we replace $f_1$ by $f_2$.

For a concrete  example,  let  $p_1$ and $p_2$ be the correlations
given in (\ref{cor:chsh}) and (\ref{corr:MSG}), respectively and
define $p_{1,2}$ to be  the corresponding product correlation.
Following the discussion above, to generate the correlation
$p_{1,2}$, one would need a Hilbert space of (local) dimension $8,$
and there is no way to save on resources in this case. {Note that
using this idea we can construct quantum correlations with various
input and output sizes on which our lower bound is tight.}

Also, if it happens to be the case that our lower bound witnesses
that $p_i$ is not finite-dimensional quantum for some $i \in \{ 1,
\ldots, k \}$, e.g., if $p_i$ is the example (\ref{cor:pr})} for
some $d \geq 1$, then $p_{1, \ldots, k}$ cannot be
finite-dimensional quantum either.

{\em {Relation to  Positive Semidefinite Rank (PSD-rank)}.} {As our
last example, we show that our lower bound on Hilbert space
dimension has a close relationship with lower bounds for the
PSD-rank.} {The PSD-rank of an entry-wise nonnegative $n \times m$
matrix $X$ is the smallest integer $c\ge 1$ such that there exist
${c \times c}$ positive semidefinite matrices $A_1, \ldots, A_n,
B_1, \ldots, B_m$ satisfying
\begin{equation}\label{eq:psdrank}
X_{i,j} = \Tr(A_i B_j), \textup{ for all } i,j.
\end{equation}

{Note the  resemblance of \eqref{eq:psdrank} to Condition
(\ref{condition:1}).} Now consider a Bell scenario  where
$|X|=|Y|=1$, i.e., Alice and Bob each have only one choice of
measurement. In this setting {we have that} any correlation
$p=(p(ab))$ is quantum and $\mathcal{D}(p)$ is known as the
\emph{quantum correlation complexity} of $p$ \cite{Zhang12}. In
\cite{JSWZ13} it is shown that in this special case,
$\mathcal{D}(p)$ is equal to the PSD-rank of the corresponding
correlation matrix $\sum_{a,b}p(ab)\ket{a}\bra{b}$, {where the
vectors are in the computational basis. Thus, our lower bound
specialized to the case $|X|=|Y|=1$ becomes a lower bound for
PSD-rank itself, which was first given in~\cite{LWdW14}. We point
out that lower bounding the PSD-rank is an important task in
mathematical optimization and quantum communication complexity
theory~\cite{LSR14}.

For general Bell scenarios, we note that the PSD-rank of the matrix
$\sum_{a,b,x,y} p(ab|xy) \; \ket{xa} \bra{yb} $ is a lower bound on
$\mathcal{D}(p)$, thus the lower bounds for the PSD-rank can also be
used to lower bound $\mathcal{D}(p)$. As an example, we consider the
correlation given in \eqref{corr:MSG}. When viewed as a lower bound
on $\mathcal{D}(p)$, the lower bound on the PSD-rank from
\cite{LWdW14} is equal to $2$, while {our lower  bound
\eqref{eq:lbound}} gives~4.

{\em Conclusions.---}In this work we derived a tractable lower bound
for the {minimum} dimension of a Hilbert space needed to generate {a
given} two-party quantum correlation {and gave a variety of
applications.} Since quantum correlations constitute a fundamental
concept in quantum physics and Hilbert space dimension is regarded
as an expensive and valuable resource, we hope our results will
provide new insights for studying quantum correlations and prove to
be useful for {their applications}. {As an example, our lower bound
has the feature that it is composed of very simple functions of the
probabilities $\pabxy$. This is very useful for analyzing the effect
of perturbations or uncertainty in the correlation data. Suppose two
experimentalists create their estimate $p'$ for the \emph{actual}
value of the correlation $p$. Then, they can use the lower bounds
(\ref{eqn:11}) and (\ref{eqn:12}) to get an \emph{estimate} for the
\emph{actual} dimensions of their quantum systems, {if they know
that for all} $a,b,x,y$, they have $|p(ab|xy) - p'(ab|xy)| \leq
\epsilon$, for {some small positive constant $\epsilon$}. In other
words, there is some threshold for the number of experiments needed
such that the two parties are fairly confident that the dimensions
of their quantum systems is at least $1$ less than {the value given
by our lower bounds when applied to their experimental data}. Thus,
our bound is quite robust against experimental uncertainty.}

\begin{acknowledgments} We thank Carlos A. P\'{e}rez-Delgado, Sixia
Yu, {and Elie Wolfe} for helpful discussions, and K\'aroly F. P\'al
and Tam\'as V\'ertesi for sending us numerical data. J.S. is
supported in part by NSERC Canada. A.V. and Z.W. are supported in
part by the Singapore National Research Foundation under NRF RF
Award No.~NRF-NRFF2013-13. Research at the Centre for Quantum
Technologies is partially funded through the Tier 3 Grant ``Random
numbers from quantum processes,'' (MOE2012-T3-1-009).
\end{acknowledgments}

\end{document}